# Intelligent Irrigation System Based on Arduino

**Authors:** C. Gilarranz [(1)], S. Altares [(1)], M. Loizu [(2)]

[(1)] *Universidad Politécnica de Madrid, Ciudad Universitaria s/n, Madrid 28040, España.*

[(2)] *Universitat Politècnica de Catalunya, Castelldefels s/n, Barcelona 08860, España.*

## ABSTRACT

This paper explains how to build an intelligent irrigation system using Arduino (a micro-controller) and many devices (humidity, temperature, pressure and water flow sensors). Our irrigation system combines a precise method to determine water balance of soils with an automatic response to water content oscillations. Thus, it is an example of how we can perform better irrigation systems by increasing the precision of measurements but also by automating decisions.

## 1. INTRODUCTION

In the past, farmers had to deal with exhausting tasks due to the limited existence of technology. The first agriculture revolution allowed the arrival of new technologies (as tractors). This meant a drastic change for farmers, as they could be much more efficient. Nowadays, we are experiencing the third agriculture revolution, which is bound to digital revolution. The new technologies that have appeared (and those that are to arrive) are drastically transforming the way farmers produce crops.

Until recently, flood irrigation has been the most extended irrigation system. Nowadays, drips and sprinkles systems are replacing it, as they are much more efficient (90-95% and 70-75% respectively) (Maisiri, Senzanje (1)). At the present time, we dispose of two methods to determine water plants requirements. The first one has been used for a long time (it is still used) and it based on crops evapotranspiration (Allen, Pereira (2)). The second method is the one we are using in our experiment. It is based on measuring soil water content in real time (with sensors), which allows a very precise management of the irrigation system. However, the calibration of a moisture sensor is very different in the field as there are much more parameters influencing lectures (Michot, Benderitter (3)). This will vary depending on the sensor used but also on the precision expected.

In this paper, we show how to combine this precise method with automation. It allows not only a fast and comfortable way to do things but most of the time it also allows less human mistakes (having machine errors in contrast).

## 2. OBJECTIVES AND SYSTEM DESCRIPTION

The objective of this experiment is to show that the automation of a complete irrigation system is possible, increasing accuracy and efficiency in terms of water application (thanks to sensors) but also allowing a more comfortable management.

Our hydraulic system is a simplified irrigation system (normally composed by a tertiary pipe, the irrigation pipes and the drippers) as we make the experiment in a laboratory and dimensions are a restriction. The irrigation pipe we use is made of *low-density polyethylene* (LDP) and working pressure is set to 150 - 200 kPa (1,5 – 2,0 bar).

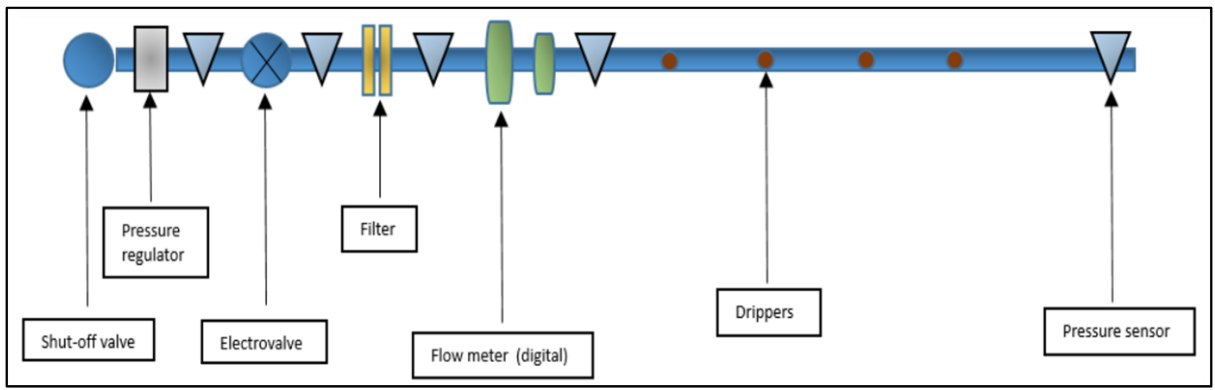

*Figure 1. System Scheme*

All components (except the shut-off valve, the pressure regulator and the first pressure sensor) have analogical and digital versions, connected in parallel (measuring the same value). Digital and analogical flow meter sensors are connected in chain, but they are close enough to give more or less the same value.

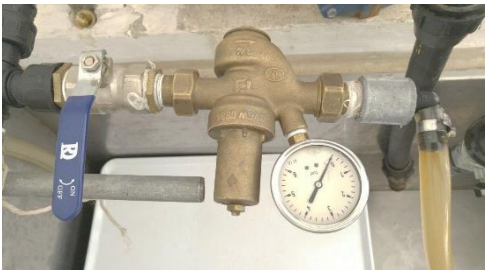

*Figure 2. Shut-off valve, pressure regulator and pressure sensor (manometer).*

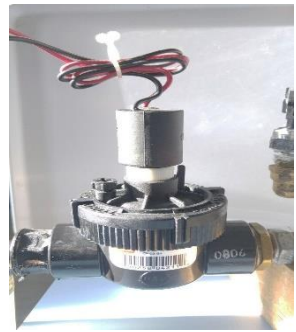

*Figure 3. Electro-valve (Solenoid valve)*

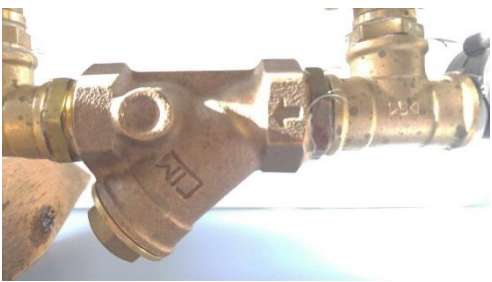

*Figure 4. Filter*

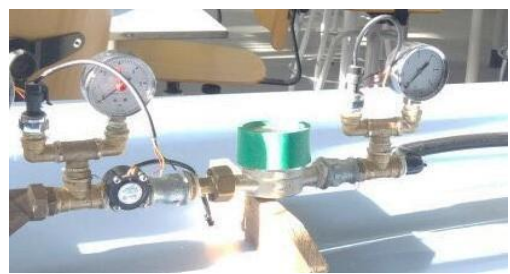

*Figure 5. Pressure sensors (digital and analogical) and flow meter sensors (digital and analogical)*

# 3. HYDRAULIC AND AGRONOMIC PARAMETERS AND CRITERIONS

To accomplish our objective, we need to set which parameters are going to be measured. First of all, we want to be sure the system is correctly working (Arviza (4)). To do that, we measure pressure at many points of the pipe:

1. The first pressure sensor (located just after the pressure regulator) will confirm if water is entering the system or not.
2. The second and third measures are done right before and after the filter, respectively. This is necessary to know if the filter is saturated. If the pressure difference between both measurements is too high, then the filter needs to be changed.
3. The fourth measurement is taken by flow meters, which is important as water flow is the main parameter in our system (it's what drippers will finally bring to crops).
4. The last two measurements are done at the beginning and at the end of the irrigation pipe. They measure what we call "initial pressure" and "final pressure", respectively. This lets us know if drippers are clogged by measuring the difference between both values: if that difference is equal to zero, it means there is no water discharge and so that drippers are clogged.

Once we have established how we measure the correct operation of the system, we need to know when to activate it (by opening the electro-valve). To do so, we measure soil water content (with a soil moisture sensor). When soil water content is below 50%, the system has to get activated. When soil water content is near 100% water flow has to be shut-off.

All this measurements are taken analogically and digitally (except for the first one which consist only in one analogical sensor, a manometer). The digital measurements need to be calibrated in order to obtain the same value that the analogical ones give us. Here is an example of calibration:

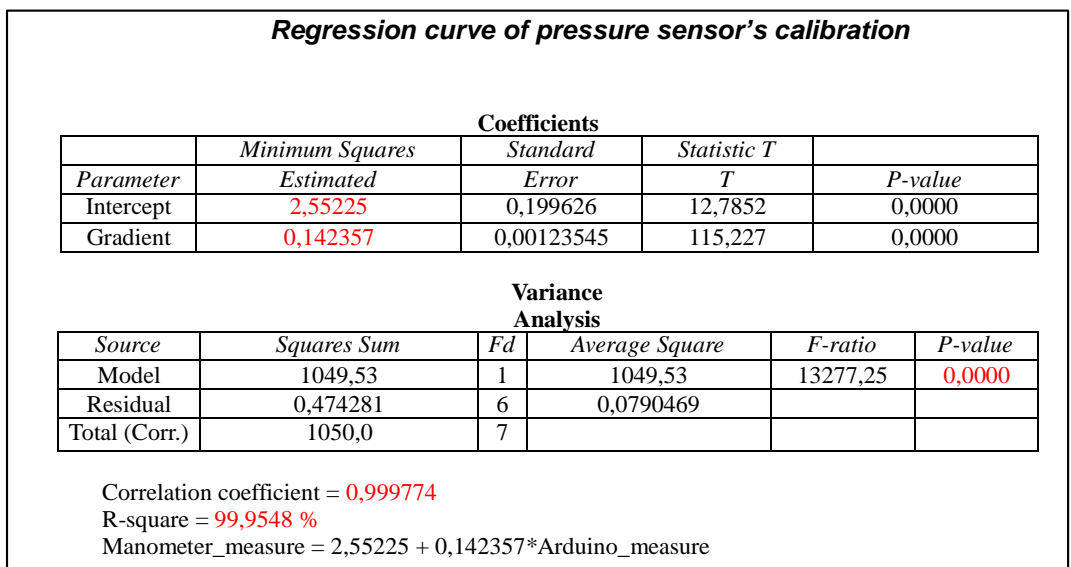

### Regression curve of pressure sensor's calibration

**Coefficients**

| Parameter | Minimum Squares Estimated | Standard Error | Statistic T | P-value |
|---|---|---|---|---|
| Intercept | 2,55225 | 0,199626 | 12,7852 | 0,0000 |
| Gradient | 0,142357 | 0,00123545 | 115,227 | 0,0000 |

**Variance Analysis**

| Source | Squares Sum | Fd | Average Square | F-ratio | P-value |
|---|---|---|---|---|---|
| Model | 1049,53 | 1 | 1049,53 | 13277,25 | 0,0000 |
| Residual | 0,474281 | 6 | 0,0790469 | | |
| Total (Corr.) | 1050,0 | 7 | | | |

Correlation coefficient = 0,999774
R-square = 99,9548 %
Manometer_measure = 2,55225 + 0,142357*Arduino_measure

*Figure 6. Regression curve of pressure sensor's calibration*

## 4. ELECTRIC AND ELECTRONIC SYSTEM DESCRIPTION

All the automation system is based on Arduino. Arduino is a company that provides many kinds of devices, as source of open hardware and software. The variety of applications are enormous (Barrett (5)). The most important is the microcontroller (in our case, Arduino UNO). It also provides a software that allows to program the Arduino microcontroller. Sensors that we use are suitable with Arduino UNO. This microcontroller has many pins (distinguishable by their type of connection: analogical or digital). Pins are the communication path between sensors and the microcontroller.

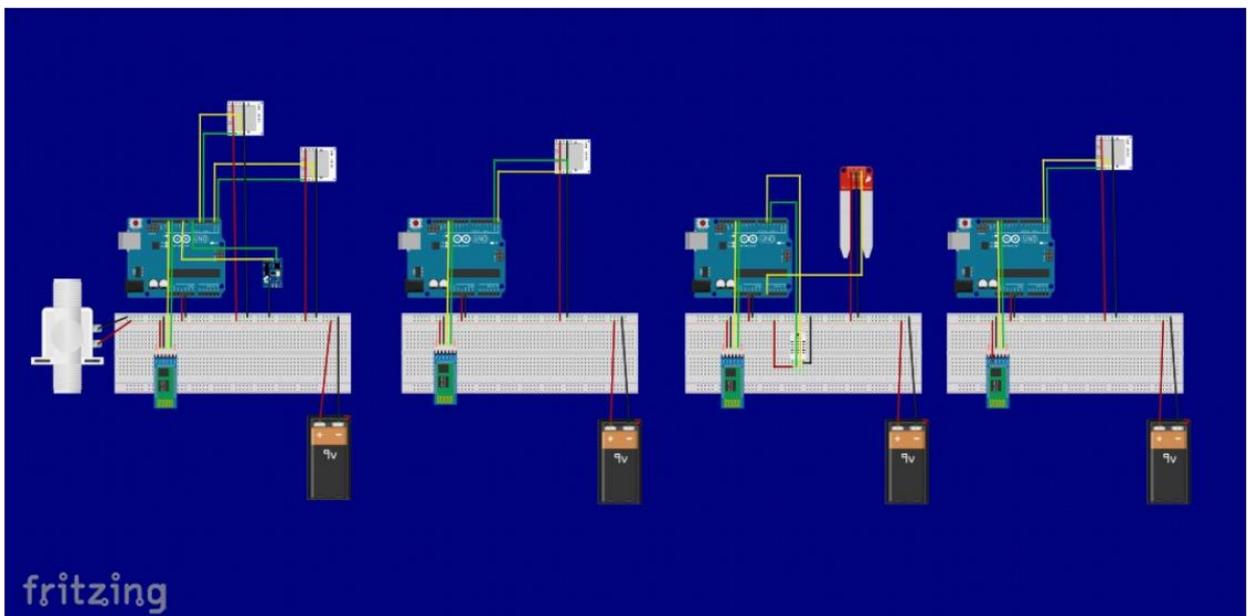

*Figure 7. Electric and electronic system scheme*

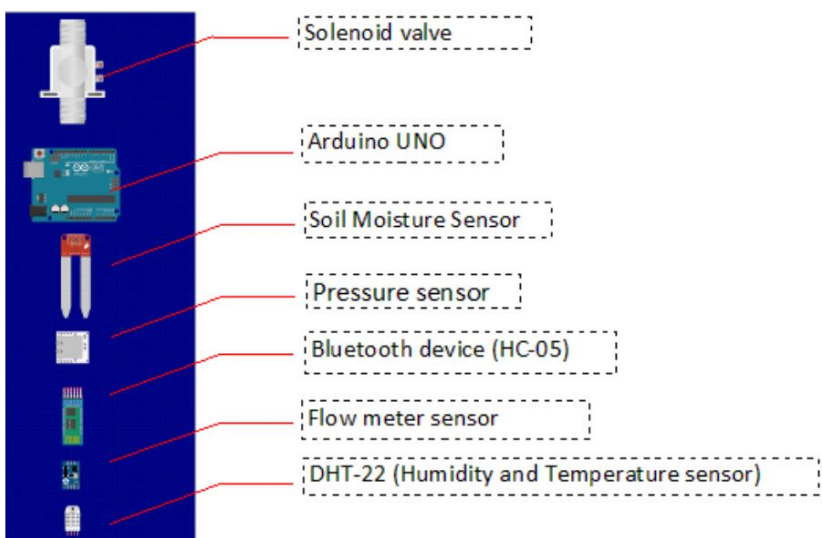

*Figure 8. Key of the scheme*

As we can see, we have four different microcontrollers. We could use only one, by connecting it to all the sensors, but this would require long wires, which would be very unpractical in a real irrigation system. That's why we decided to use Bluetooth devices (in our case, HC-05 model), which allows communication between microcontrollers (Bravo-Pérez, Redondo-Aycardi (6)). The decision making has to rely on only one microcontroller (the "master") which will receive data from the others (the "slaves"). However the master microcontroller can also receive data directly from sensors connected to it (which is our case).

Each microcontroller is placed near pressure sensors in order to reduce the amount of wire used. Then, they are connected to all nearby sensors, as showed in Figure 7. Soil moisture and air temperature and humidity sensors are placed in the irrigation zone (near the drippers) as it is where plants are located.

Data retrieved from sensors will be integrated by microcontrollers. Then, all slaves microcontrollers will send the data to the master microcontroller, which will transform it according to calibrations. Here is an example of the code used for the master microcontroller:

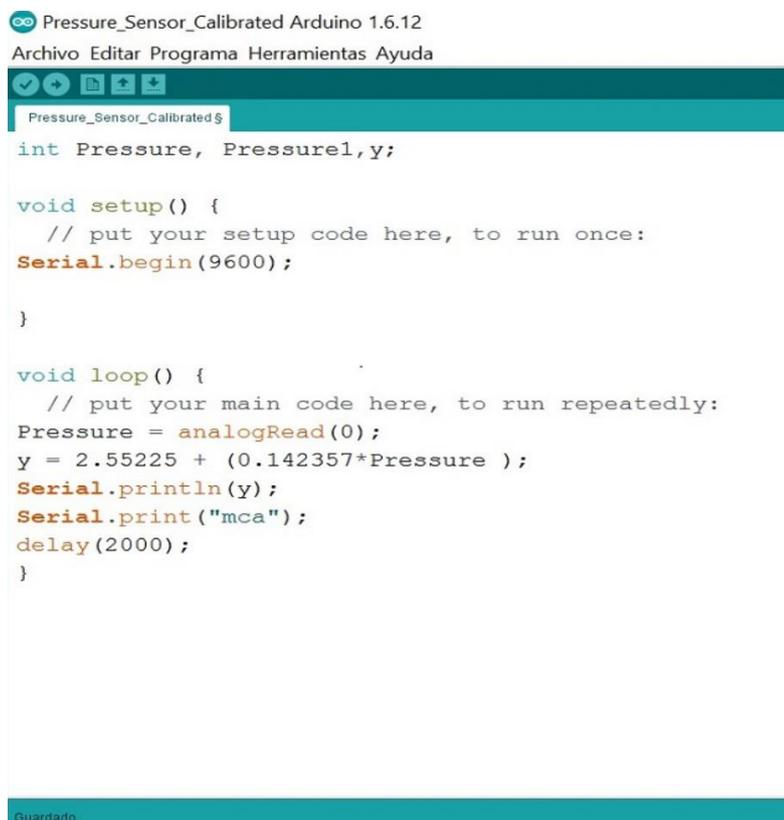

When the master microcontroller has received and transformed all the data, it will evaluate that information and compare it to the restrictions exposed initially. Then it will send a signal to the solenoid valve to get it opened or closed. Concerning technical problems (as drippers clogging, filter saturation or others), some other devices can be implemented in order to warn the person in charge: a Display (which would show a text), a buzzer (producing a loud noise) or a LED.

## 5. CONCLUSIONS

We would like to conclude by saying that this system works successfully. We have shown that there are other ways to manage crops and ornamental plants, which allows us to be much more efficient in terms of time, money and water waste. It seems clear that this kind of prototype can be extrapolated to a complete irrigation system. Thus, we could expect to see this kind of systems appear increasingly in agriculture. This is why farmers and agricultural companies should begin applying this kind of technology in their ordinary activities.